\documentclass[11pt, a4paper]{article}
\usepackage{jheppub}
\pdfoutput=1

\begin{document}

\newcommand{\be}{\begin{equation}}
\newcommand{\ee}{\end{equation}}
\newcommand{\bea}{\begin{eqnarray}}
\newcommand{\eea}{\end{eqnarray}}
\newcommand{\ba}{\begin{array}}
\newcommand{\ea}{\end{array}}
\newcommand{\pt}{\partial}
\newcommand{\nn}{\nonumber}
\def\tr{\mathop{\rm tr}\nolimits}
\def\Tr{\mathop{\rm Tr}\nolimits}
\newcommand{\dsf}{\displaystyle\frac}
\newcommand{\ds}{\displaystyle}
\newcommand{\pz}{\partial_z}
\newcommand{\dz}{\Delta_z}
\newcommand{\dzf}{\Delta_{z,4}}
\newcommand{\ypp}{f^{\prime\prime}}
\newcommand{\yp}{f^{\prime}}
\newcommand{\Zpp}{Z^{\prime\prime}}
\newcommand{\Zp}{Z^{\prime}}
\newcommand{\dzp}{\Delta^+}
\newcommand{\dzm}{\Delta^-}
\newcommand{\dzpt}{\tilde\Delta^+}
\newcommand{\dzmt}{\tilde\Delta^-}
\newcommand{\dzpm}{\Delta^\pm}
\newcommand{\dzmp}{\Delta^\mp}
\newcommand{\mn}{p}
\newcommand{\mnm}{p_m}
\newcommand{\tM}{\tilde M}
\newcommand{\ts}{\tilde\sigma}
\newcommand{\tp}{\tilde\mn}


\def\a{\alpha}
\def\b{\beta}
\def\c{\gamma}
\def\d{\delta}
\def\e{\epsilon}
\def\ve{\varepsilon}
\def\f{\phi}
\def\v{\varphi}
\def\g{\gamma}
\def\h{\eta}
\def\i{\iota}
\def\j{\psi}
\def\k{\kappa}
\def\l{\lambda}
\def\m{\mu}
\def\n{\nu}
\def\o{\omega}
\def\p{\pi}
\def\th{\theta}
\def\r{\rho}
\def\s{\sigma}
\def\t{\tau}
\def\u{\upsilon}
\def\x{\xi}
\def\z{\zeta}
\def\D{\Delta}
\def\F{\Phi}
\def\G{\Gamma}
\def\L{\Lambda}
\def\O{\Omega}
\def\P{\Pi}
\def\Q{\Theta}
\def\S{\Sigma}
\def\U{\Upsilon}
\def\X{\Xi}
\def\delb{\bar\del}

\title{The physical mechanism of AdS instability 
and Holographic Thermalization.}

\author[a]{Sang-Jin Sin}
\affiliation[a]{Department of Physics, Hanyang Univ.  Seoul 133-791, Korea}
\emailAdd{sjsin@hanyang.ac.kr}

\abstract{
Gravitational falling in AdS has two characteristic properties \cite{ohsin}: i) A thick shell  becomes a thin shell. ii) Any shape become spherical. 
Such focusing character of AdS, for the collapse of dusts, leads to the rapid thermalization mechanism in strongly interacting system. 
  For the collapse of a wave, it explains the  
cascade  of energy to UV through   repeated bounces, 
which has been extensively discussed in recent numerical works.
Therefore the focusing is the physical  mechanism of instability of AdS.
Such sharp contrast between the dust and wave in collapse, together with the experimental observation of rapid thermalization, suggest that the initial condition of created particles in RHIC is in a state with random character rather than a coherent one.
Two time scales, one for thermalization and the other for hydro-nization are defined and calculated in terms of the total mass density and energy distribution of the initial particles. 
We find $t_{th}\sim (1-c_1/E^2)^{1/2}/T $ so that softer modes thermalize earlier.  However, for hydro-nization, 
 $t_{hyd} \sim 1/E^{2/3}T^{1/3}$ therefore harder modes come earlier. 
 We also show that near horizon limit of Dp brane solutions  have similar focusing effect which is enough to guarantee the early thermalization.
}

\maketitle

\section{Introduction }
\label{sec:Introduction}
 
The thermodynamics is extremely useful to describe the
nature, although a true  equilibriated  state is very rare. It can be attributed to the fact that thermalization is a very rapid process. Nature seems to know 
how to arrive at the equilibrium without overshooting, which is more effective if the system is strongly interacting.  
One  example of extremely rapid thermalization\cite{Heinz:2001xi}
is given by Relativistic Heavy Ion Collision (RHIC)  experiment: the fireball seems  to reach equilibrium in 1fm/c, which is a time
for gold ions  to pass each other. Certainly this is due to the strongly interacting nature of the quark gluon plasma although the precise mechanism has not been understood, yet.
Since understanding  this phenomena is beyond perturbative field
theory, it is natural to ask if a dual formulation can shed any light on it.  

According to the gauge/gravity duality \cite{Maldacena:1997re},
thermalized state is dual to black hole geometry, and therefore 
the  thermalization process is dual to the black hole formation 
in the dual picture. 
Some time ago, in \cite{Shuryak:2005ia}, mapping the entire
process of RHIC experiment to the dual gravity language has been tried.
Since expansion of the RHIC fireball should be dual to the falling
of the matter in the dual gravity, black hole formation from the 
the collapsing matter seems to be a natural candidate as a dual process of fireball equilibration. 

Since there have been many works on this issue, we briefly mention  difficulties of existing models.  With spherical collapse of scalar wave in global anti-de Ditter space (AdS)   
\cite{Danielsson:1999fa, Balasubramanian:2011ur}, or  
\cite{minwalla} with homogeneous null hyperplane source in POincare patch, although  both of them have mathematical success    due to  their simplicity,  the initial  conditions assumed there is too fine tuned to be connected to a real process of thermalization: to an observer outside the shell,   the spherical {\it initial} configuration is equivalent to preparing an already equilibrated system  by the Birkoff theorem.  
Another useful work is the colliding shock wave model\cite{chesler}. Its problem is that the initial condition is set BEFORE the collision where gravity dual is not relevant. 
The rapid particle creation is the key  reason why the fire-ball of RHIC acts as a strongly interacting system: it converts  almost all initial kinetic energy into the mass of the created particles such that only $0.1\%$ of kinetic energy play the role of the temperature. The gravity dual is responsible only in this strongly interacting regime which is realized only AFTER the collision. 
Late time expansion of hydronized  QGP was  obtained by 
using the falling horizon model in \cite{Janik:2005zt}. 
There, the goal was the embedding the adiabatic cooling into gravity formalism rather than the mechanism  to achieve such quasi-equilibrium configuration.   
So it is still not clear  whether a black hole will be formed starting with  generic  initial configurations. 

In flat space,
matter without dissipation mechanism can not make a black hole.  
Moreover, recent studies on scalar field collapse 
\cite{Bizon:2011gg,Dias:2011ss,Garfinkle:2011hm,Buchel:2012uh,Bantilan:2012vu}
show  that even for the spherical shell in AdS space, black hole is formed only after many repeated  reflections from the boundary, which
is certainly not the dual of `thermalization
in one passing time'.  The natural explanation for the experiment
should be such that the black hole is formed in one falling time without any oscillation  for any non-spherical/non-homogenous initial configuration in the global-AdS/Poincare-patch.

In the previous paper \cite{ohsin}, we addressed this issue using the special property of the AdS space: any geodesic has the same period and  therefore particles arrive at the center simultaneously regardless
 of their initial position if they start from zero velocity. As a result, a shell of dust particles
  with arbitrary shape becomes spherical as
 it falls and it forms black hole when the last particle pass the apparent horizon. It takes a time less than one falling time.
 Another consequence is that thickness of the initial shell becomes thinner as it falls. See figure 1.
 
%
%
\begin{figure}[t]
\centering
\includegraphics[width=8cm]{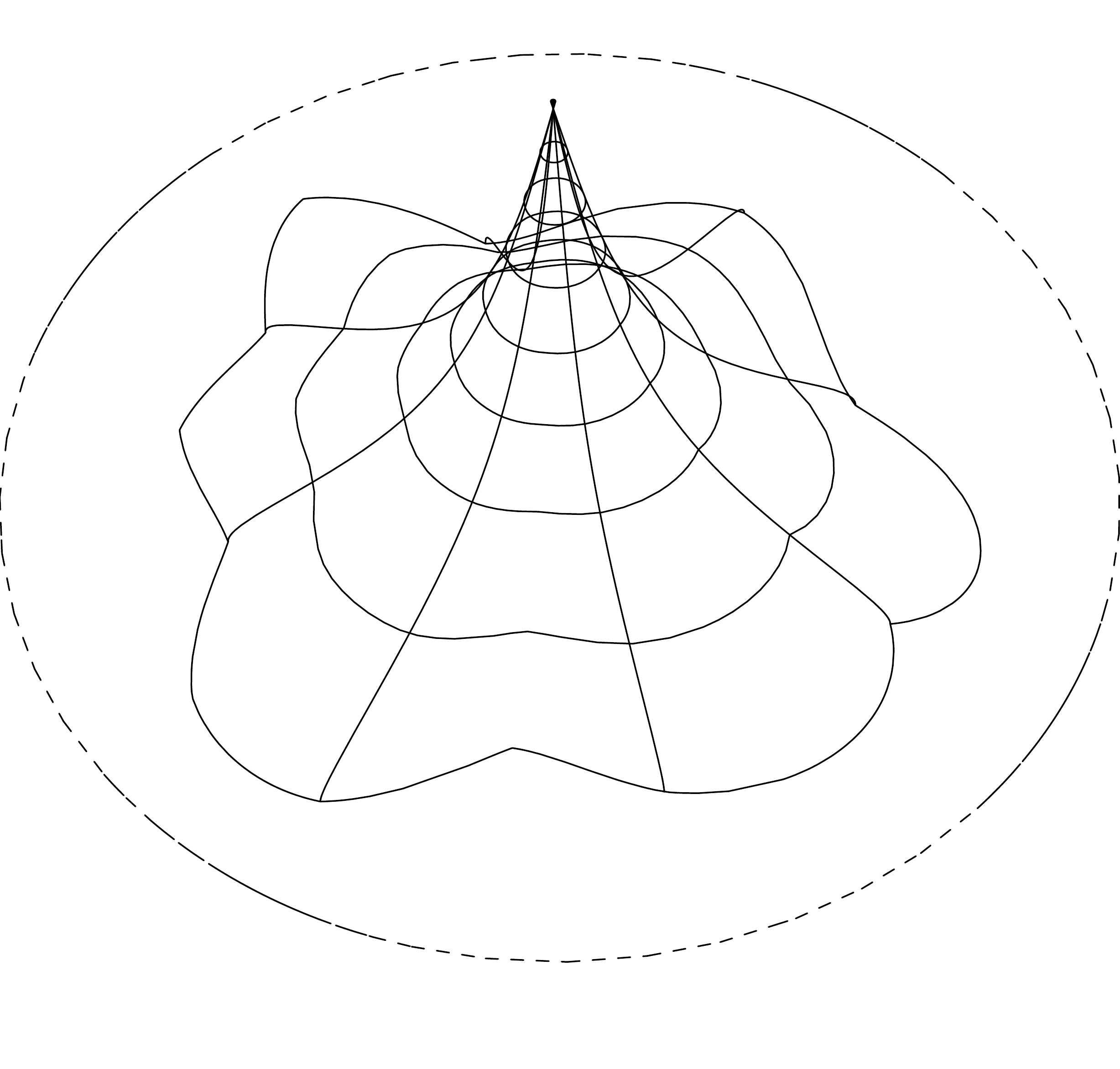}
\caption{Falling of a shell in global AdS represented as a cylinder whose vertical axis is time direction. 
Focusing effect  of AdS gravity results in two remarkable phenomena:  i) A thick shell becomes thins shell. ii)An ugly shell becomes perfect sphere. The first leads to cascade-to-UV/black hole formation, and the second leads to the isotropization/hydronization.  
 }
\label{fig:focusing}
\end{figure}

In a Poincare patch, focusing phenomena are robust even in the case we include inter-particle interactions and initial velocity in any non-radial direction.  At the moment of creation,   zero  radial velocities  should be assumed  for the holographic images of 5000 created particles since there is no reason why  they should be moving  along holographic direction at the moment of creation. We   call this  as focusing mechanism of dual gravities.

Here, we will first discuss the wave collapse and suggest that focusing is the physical mechanism of instability of AdS.
Then,  we will discuss more quantitative prediction of the focusing mechanism: we will conceptually distinguish thermalization time  and hydronization time and then 
and calculate them.   We will also show that near horizon limit of Dp brane solutions  have similar focusing effect, indicating that 
the focusing mechanism is the universal feature of any gravity dual.    
At the end, we will summarize by listing about 15 reasons why focusing is the mechanism of the effective thermalization in gravity dual picture.  

\section{Mechanism of energy cascade to UV in wave collapse. }\label{sec:wave}
 What will  happen  to the collapse of wave rather than  dust particles? 
 First, one should be notice that it costs energy to  localize a
wave packet in a small region,  which is the origin of the uncertainly principle of wave mechanics. 
Such dispersive nature is also responsible for the stability of our material world not  collapsing  down to a neutron star and also for the stability of boson star \cite{sin}. Gravity can  confine a wave
packet within   the Schwartzshild radius   $r_s$ only if the initial
configuration is thin and spherical enough, equivalently only if $r_s$ is large enough. 
For   generic configuration  gravitational potential energy is not enough to provide such localization.
Therefore
generic initial wave configuration should bounce back after initial collapse and then fall again. 
Since nothing can escape from the AdS, which is like a box, the
same process repeats again and again. 

Why a black hole is
formed eventually after enough number of bounces? We can understand this if we visualize a shell-like  wave configuration as a collection of infinite number of particles connected by springs. 
Each time the shell falls, the thickness of the shell
decreases due to the focusing mechanism of the AdS. 
When it bounces back, the decreased thickness does not go back to its initial state due to the attractive inter-particle interaction.
After enough number of bounces, the shell becomes thin  and
spherical enough so that entire wave configuration can be inside its Schwarzshild radius. 

In momentum space, these geometric progress   
has an interesting interpretation. Whatever is the initial
configuration of the wave packet, as times goes on, the angular distribution will be
shifted towards $l=0$ to become spherical  while the radial distribution will be shifted to
larger momentum region to make the shell  thinner. That is, cascade of amplitudes to UV regime is derived  by the focusing mechanism of AdS space. We believe that this is what is happening in  recent numerical works \cite{Garfinkle:2011hm, Buchel:2012uh, Bantilan:2012vu} with initial spherical shell.  
If one starts with initially non-spherical shell, one would also  observe   a cascade of the angular distribution towards  zero angular momentum as well as the cascade to UV.
 
Summarizing, in global AdS, the collapse of  arbitrary shape of dust shell forms a black hole at once,  while  that  of wave shell makes black hole after enough number of oscillations. In both cases, focusing mechamism is the underlying mechanism for the eventual black hole formation. 
We believe that the focusing is the physical mechanism of the instability of AdS \cite{Bizon:2011gg,Dias:2011ss}: If we dump anything to AdS, it is unstable to become a black hole, although the time depends on the nature of the matter. 

Since what is observed in heavy 
ion collision is `thermalization at once', the initial state of
fireball created in heavy ion collision should be considered as an  incoherent dust-like 
state  rather than  a classical wave configuration,
which is dual to particles in a condensed  state.

\section{Focusing in other dual gravities.}
\label{sec:3}

Now we come to the one of the main question: Is the focusing mechanism is special and  working only for AdS space, or generic to a large class of gravities? 
If the answer is the former, focusing is not very useful 
as a mechanism of  thermalization, because we do not know whether the gravity dual of the given system is
precisely the AdS. Here, we will show that the field theory limit of all Dp branes solutions have focusing mechanism
if $p\le 5$. 

Particles interacting via gluon exchange are mapped into non-interacting particles moving 
in the background AdS space. Only the residual non-gluonic
interaction should be handled by inter-particle interaction.
Therefore we consider only non-interacting particles, moving in a gravitational background given by
 \be ds^2=-g_{tt}dt^2+ g_{rr}dr^2 + g_{ii}dx^i dx^i .\ee
The equation of motion is given by the action 
\be S=- m\int \sqrt{-
g_{\mu\nu} \dot{x^\mu} \dot{x^\nu}} dt, \;{\rm with} \;\; {\dot
x}=\frac{dx}{dt} . \ee
 We set $R=1, c=1$.
The radial motion can be first integrated to give the energy
conservation:  
\be
 \frac{m g_{tt}}{\sqrt{g_{tt} -g_{rr}{\dot r}^2 }}=E.
 \label{1st-int}
 \ee
 If the motion starts with zero radial velocity from the
 initial radial position $r_0$, then
 \be E=m \sqrt{g_{tt}(r_0)}, \label{E-r0}
 \ee
which can be used as a dictionary between the conserved energy E and the
initial radial position. 
Namely we can assign the initial position $r_0$ for a particle whose
energy is $E$ such that they are related by eq.(\ref{E-r0}).

 Introducing  $v_c=\sqrt{  {g_{tt}(r_0)}/{(1+g_{tt}(r_0) )}}$,
we have $ E =  {m}/{\sqrt{1-v_c^2}}.$
Therefore $v_c$ can be interpreted as radial velocity when it arrives at
the center.
  The equation of motion can be formally integrated: 
 \be
t= \int_r^{r_0}\frac{dr}{ \sqrt{ {g_{tt}(r)}/{g_{rr}(r)}\cdot
(1-g_{tt}(r)/\epsilon^2)}}, \label{time}
 \ee
where the $\epsilon=E/m$ is the energy per unit mass.

For the global AdS metric, the period of the
motion is 
$T_{fall}= \frac{\pi  }{2} \frac{R }{c}$, which is  
independent of the initial position. This is in fact a unique property of AdS, which is not shared by any other metric known to us. If thermalization is really depends on this exact synchronization, the focusing mechanism would not be a universal
one and we have a less chance to explain the RHIC experiment 
in terms of the focusing mechanism. 

Fortunately, it turns out that the exact synchronization in finite time is not really necessary for thermalization.  
 For the Poincare patch of the AdS, 
 $g_{tt}=r^2=1/g_{rr}$  and the falling time  is infinite. 
 However, as we can see from the expansion 
 \cite{Shuryak:2005ia},
\be
r=  \frac \epsilon{ \sqrt{1+ (\epsilon t)^2}} =\frac 1 t 
-\frac{1}{2\epsilon^2}\frac 1{ t^2} +   {\cal O}(1/t^4).
\ee
 the leading term  is independent of  initial condition and the subleading terms rapidly vanish as time
goes on. Therefore  all the particles will be inside the apparent horizon within finite time and we conclude that  focusing mechanism still  works for the AdS with flat boundary. 

This mechanism does not request arrival at the center simultaneously for many particles. 
Infinite falling time 
with bounding gravitation potential together with the finite 
would-be-horizon radius seem to be all we need.   
To support  this conjecture, we now show that above mentioned 
 property holds for  near horion geometry 
of other $D_p$ branes.
Let's start with D4 case where 
  $g_{tt}=r^{3/2}=1/g_{rr}$. 
  We can show that 
  \be
  r(t)=4/t^2 - 8a/(\epsilon t)^3 + \cdots,  
  \ee
where $ a=2 \sqrt{\pi} \Gamma(2/3)/\Gamma(1/6) \simeq .86237$.
 So, only after $t\to \infty$ $r(t)\to 0$ meaning that falling time is infinite and the leading term is independent of the initial condition. Therefore 
 falling in the $D4$ 
is qualitatively the same as  that in  the Poincare patch of AdS.

For general $D_p$ brane case,  
\be
r(t)=\left( \beta /{t} \right)^{\beta} +  \cdots, ~~ with ~~ \beta= \frac{2}{5-p},  ~~\hbox{for } p < 5.
\ee
The difference in the initial condition will decay away with power law. 
Using the hyper-geometric function, we can express the exact solution: 
$$
t=\frac{1}{\alpha-1}\left[ 
 \frac{1}{r^{\alpha-1}} \;_2\! F_1\left(\frac 1\alpha -1 , \frac 1 2 ,\frac 1\alpha; \frac{r^\alpha}{r_0^\alpha}\right)  -\frac{\Gamma(\frac{1}{\alpha})  \Gamma(\frac{1}{2})}{
\Gamma(\frac{1}{\alpha} -\frac 1 2)} \frac{1}{r_0^{\alpha-1}} \right] ,
$$ where $\alpha=\frac{7-p}2$ and $r_0^\alpha=\epsilon^2$.
For p=5, we can find a simple solution 
\be 
r(t)= r_0/\sinh^2(t/2) , 
\ee
so that the difference in the initial positions will be exponentially washed out: $\Delta r \sim 4\Delta r_0 \cdot e^{-t}$  for late time. Notice that $r_0=\epsilon^2$ for p=5 with eq.(\ref{E-r0}). 

Technically, the origin of the focusing mechanism in these background is the behavior of metric 
$g_{tt}/g_{rr}\sim r^\alpha$ with $\alpha \ge 1$, which in turn can be attributed to taking 
 the Maldacena limit from the Dp geometry: Had we keep the 1 in $H_p=1+c/r^{7-p}$, 
we would not have  focusing mechanism. So the 
focusing mechanism  is the property of neck region of original $D_p$ geometry.

\section{Conserved Energy v.s  Radial position : AdS/CFT dictionary}
So far we mainly discussed the radial motion. 
In the presence of the  velocity in the boundary direction, 
we have difficulty in the global AdS since that will create angular momentum and therefore build up centrifugal barrier, which 
will forbid falling to form the black hole formation. 
To avoid this, we examine the effect of the initial horizontal velocity  in the Poincare patch where the equation of motion is given by 
\be
 \frac{mr^2}{\sqrt{r^2 (1-{\dot x}^2)-{\dot r}^2/r^2}}=E, \\
  \frac{mr^2 {\dot x}}{\sqrt{r^2 (1-{\dot x}^2)-{\dot r}^2/r^2}}=p, 
 \label{1st-int4}. 
 \ee
 At the moment of creation ${\dot r}=0$ although ${\ddot r}\ne 0$, so that 
 \be  r_0\cdot \frac{m}{\sqrt{1-v^2}} =E  , \;\;   r_0 \cdot \frac{mv}{\sqrt{1-v^2}} = p \label{dic}
 \ee
 The right hand sides of above equations are conserved energy and momentum 
 while the left hand sides (LHS) are product of two: initial bulk height times bulk energy/momentum. 
 If we identify the the conserved quantities as those at the 
the boundary and the energy and momentum in the LHS as bulk quantity, it is consistent with the prescription of 
Polchinski and Strassler \cite{Polchinski:2001tt}. 
According to above relation, we can attribute the boundary 
energy partly to the bulk energy and partially as bulk 
initial height such that eq.(\ref{dic}) holds. Since the bulk energy $ {m}/{\sqrt{1-v^2}}$ is always bigger than the mass $m$, $r_0$ has maximum value $E/m:=\epsilon$. The question of how much bulk kinetic energy is assigned is matter of choice and in this sense there is a {\it holographic gauge choice}. 
We choose the gauge where $r_0$ is maximum and $v=0$. Therefore in our gauge, all the  energy
of a created particle is attributed to the height of its  bulk-image. This is a justification postulated in the previous work \cite{ohsin}.

 At the boundary, the thermalization is complicated process of strongly interacting particles. But in the bulk, 
the particles `interact' only with the background metric, namely  they are just free falling many-body system. This is the simplicity obtained from the dual gravity formalism: 

 \section{Time for Thermalization and Hydro-nization}
How long does it take to thermalize? 
 A natural definition for the thermalization  is  the time at which  the last particle   in the process  passes the apparent horizon $r_H$, which is determined by the total energy inside the system:  
\be
r^4_H= \frac{c_2}{V} \sum_i E_i,  
\ee
where $c_2$ is a constant proportional to the Newton's constant $G_N$, $V$ is the volume of the 3-space  and $E_i$ is the energy of i-th particle.  
If  the system has two well separated parts in energy distribution, we can model   it  as  two well separated  shells. It is easy to prove that radial motion of particle with higher initial position  will  catch up the lower particle but never overtakes. 
Therefore, we can conclude that soft modes thermalize first and then hard modes thermalize on top of the former. 

  The thermalization time is simply given by: 
\be
t_{TH}=\sqrt{\frac{1}{r_H^2} -\frac{1}{r_0^2}} =\frac{1}{\pi T}\cdot \sqrt{1- \frac{(\pi T m  )^2}{E^2}}.
\ee
 where $r_H,$ and $T$ are the radius of the `would-be-horizon' and its corresponding temperature which will be reached by the system after thermalization. $r_0$ is the initial height of the particle with highest energy.
If we consider the  initial energy distribution of particles in the system such that it is not concentrated on specific energy scale,  then the highest energy scale is much bigger than the temperature.  If $r_H <<r_0$ the thermalization time is given by the inverse temperature,
 \be
 t_{Th} \simeq \frac{1}{r_H}=\frac{1}{\pi T}, \label{th-time}
 \ee
Otherwise, thermalization time should be less than this. 

Now  look at the falling of two particles whose initial heights 
are order of magnitude different. Although there is no passing, particles with higher radial position almost catch up the lower particles within half of the thermalization time. Such phenomena strongly suggests that approximate isotropization happens much before the actual thermalization. See figure 2. 
  
\begin{figure}[t]
\centering 
\includegraphics[width=8cm]{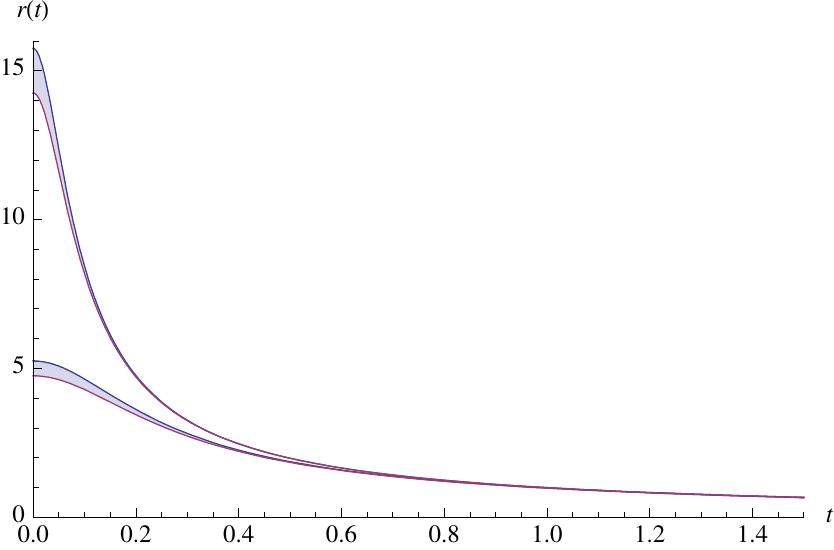}
\includegraphics[width=7cm]{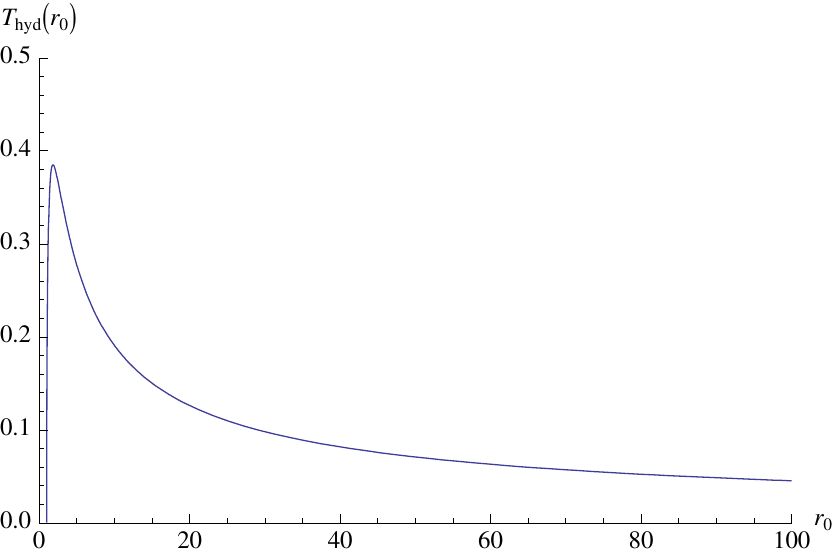}
\caption{Left: $r(t)$ v.s $t$:  Falling of two shells with different energy scales. Each shell has thickness of 10 $\%$ of its average energy. 
The higher the average position, the sooner the thickness reduces; 
Right: Hydro-nization time as function of initial height of the shell. The higher is the energy,  the sooner it becomes isotropic.  }
\label{fig:hydro}
\end{figure}

To an external observer, after rapid process of isotropization, 
not much change will happen although the shell is still falling. Only residual process of isotropization is under progress. 
We can identify such stage as hydronized state. This is the regime where  the metric is still time dependent but hydrodynamics works effectively.
One can quantify the hydro-nization time $t_{hy}$ as follows: 
Take a shell of thickness $\Delta r_0:=r_{0,max} - r_{0,min}$ which is, say,  10$\%$ of its average height $r_0$. 
This amounts to taking the particles whose initial  energy distribution width is $\Delta E/E=10\%$ around its average value $E$.
We can say that the system is hydro-nized after time $t_{hy}$, if the maximum difference of radial positions in the shell is within $10\%$  of the horizon radius, that is if 
$\Delta r(t_{hy}) /r_H \le 0.1$ which is equivalent to 
$\Delta E(t_{hy})/T \le 0.1$, which can be inverted to give  
 \be 
 t_{hy}=\frac1{r_0}\sqrt{\Big(\frac{r_0}{r_H}\Big)^{2/3} -1}
 =\frac{m}{E} \sqrt{\Big(\frac{E}{\pi Tm}\Big)^{2/3}-1} .
 \ee
 When initial energy per mass  of the shell is much bigger than the final temperature, 
 we can get 
 \be
 t_{hy}=\frac{(m/\pi^2)^{1/3}}{E^{1/3} T^{2/3}}. 
 \ee 
It says that the hydro-nization  is  faster for higher initial energy. 
Therefore hard modes hydro-nize first and then soft modes follows.

\section{Summary and Conclusion}

In this paper we answer to the question what is the 
physical mechanism for the instability of the AdS, namely why any thing dumped into AdS cause black hole formation. We also 
 demonstrated how to use it to characterize the thermalization and hydro-nization. Falling particles  arrive at the 
center simultaneously independent of the initial height, which causes two remarkable effects: It makes a shell of arbitrary shape   spherical and makes thick shell thin. 
We argued that these effects are not only the mechanism for the formation of the black hole in one falling time  when applied to the dust particles, but also are the physical mechanism  of the cascade to UV modes and cascade to lower angular momentum mode. Focusing in AdS is the underlying mechanism of the instability of AdS. 
Two distinguished time scales for hydro-nization and thermalization are defined and calculated. The former is   the time for  black hole formation and the latter is time for isotropization. 

Here we summarize  by listing the evidences why the FOCUSING and its two consequences are the mechanism of rapid thermalization/hydronization. 

\begin{enumerate} 
\item Falling in bulk  is necessary for the fireball to expand at boundary before thermalization. These two are dual to each other. 
\item It enforces dust shell to form the black hole within one falling time.
\item For the wave collapse, black hole forms only after  enough number of  oscillations \cite{Bizon:2011gg, Dias:2011ss, Garfinkle:2011hm, Buchel:2012uh}: We can understand the 
bouncing mechanism from the wave nature:   as the field
 configuration collapses gravitationally, uncertainty principle activates the kinetic term for the localized wave packet, which generates pressure and causes a bounce. 
\item The eventual formation of black hole  for the collapsing wave can be understood: each time it fall, it becomes thinner, which enforces  cascade of energy to UV. When it bounces back, attractive interaction partially preserve  two consequences of the falling. 
\item Sharp contrast between dust and wave enforces us to decide the nature of the initial condition after collision. 
The created particles should be to be non-coherent dust like state rather than a wave-like state which is dual to highly correlated many body system. 
\item We only need to consider  non-interacting particles since gluon exchange is transformed into geometric background.  
\item In poincare patch, initial non-radial velocity  does not distroy the focusing effect. 
\item Precise dictionary exist between bulk radial position and boundary energy such that we need to consider only radial falling. 
\item   Attractive residual interaction  does not distroy the focusing effect in Poincare patch. 
\item It allows quantitative discussion for time scales of Thermalization as black hole formation time.  
\item It allows quantitative discussion for time scales of  hydronization  as isotropization time.  
\item Focusing generates entropy creation and ir-reversibility by the horizon formation. 
\item The mechanism working for global AdS (the exact syncronization) is a unique property of 
AdS and it is not shared by other geometry.   
\item The mechanism working for Poincare patch of AdS 
is universal to all gravity duals, that is, infinite falling time  of near horizon geometry (of Dp branes) allows all known gravity dual to have focusing effect. 
\item Assuming a spherical symmetry or setting initial condition 
before the particle creation can be a serious problem of a model for RHIC collision. 
\end{enumerate}
 
It would be very interesting if the intuition obtained in this paper 
can be utilized to understand the cascade \cite{Bizon:2011gg, Dias:2011ss, Garfinkle:2011hm, Buchel:2012uh, Bantilan:2012vu, Adams:2012pj} and inverse-cascade \cite{Carrasco:2012nf,Adams:2013vsa, Green:2013zba, 2dinvcascade} of the energy of the holographic fluids . 
While the cascade  of 3+1 dimension looks natural from our point of view,
for inverse cascade of 2+1 dimension seems to request other idea. 
Also it is necessary to examine other backgrounds to see the universality of the focus mechanism for dual gravity. 

In this paper  we restricted ourselves in the extremal backgrounds. It would be interesting to consider the 
non-extremal backgounds. 
For confining backgrounds, the geometry cap off in the IR region and we  have a natural IR cut-off. 
In such case, some particles will bounce from the core boundary before others pass the horizon radius.   
Therefore naively it is expected to be impossible to reach thermalization at once. However, numerical experiment shows that confining metric for D3 has strong  focusing mechanism: 
The falling time from the initial height to the core boundary is almost the same and saturates to a fixed value  as its initial height increases. It may not form a black brane but may give 
the hydronization very quickly. Other possibility is that 
  if  the total mass or mass density  is large enough such that its apparent horizon radius is larger than  the core radius $r_{KK}$, falling particles might form a black brane 
after enough number of bounces. 
If the particles are falling in the background of black hole metric, there is no question of thermalization. But it is still interesting if non-passing property still holds in this background. 
We leave studying these issues to future works.

%
%

  \acknowledgments
 I'd like to thanks  
 Y.Seo, S.Seki, I. Takaaki, Y. Zhou for discussions and    
 H.Bae for drawing figures. I also appreciate questions and comments of  S. Das, J.Erdmenger, G. Horowitz, G. Policastro, G. Semenoff and T. Takayanagi. I'd like to thank APCTP, BIRS, MPI, Newton institute and YITP for the hospitality during the workshops on  holography.  
This work was supported by Mid-career Researcher Program through NRF grant No. NRF-2013R1A2A2A05004846.  It is also supported by the NRF grant through the SRC program
CQUeST with grant number 2005-0049409.


%
\end{document}